# FREE FALL FRAME TO DEMONSTRATE WEIGHTLESSNESS

RAJU BADDI

National Center for Radio Astrophysics, TIFR, Ganeshkhind P.O Bag 3, Pune University Campus, PUNE 411007, Maharashtra, INDIA;  baddi@ncra.tifr.res.in

## ABSTRACT

A simple apparatus to demonstrate free fall weightlessness is given. This can be easily constructed from readily available materials and can be used for the demonstration purpose to an audience with all the details of the apparatus visible.

## 1. INTRODUCTION

It is commonly known from physics school that freely falling bodies have zero weight(Strelkov 1987). To demonstrate this here a small electronic appartus is described which shows the loss of weight by a test object W by the glow of an LED(light emitting diode). The frame can fall freely by 3 rods/wires. On the other hand it is also possible to simply drop it on a sponge bed. The arrangement is such that during its fall if the weight of W vanishes then the LED has to glow. It is common practice to weigh something using a spring balance. Under the action of

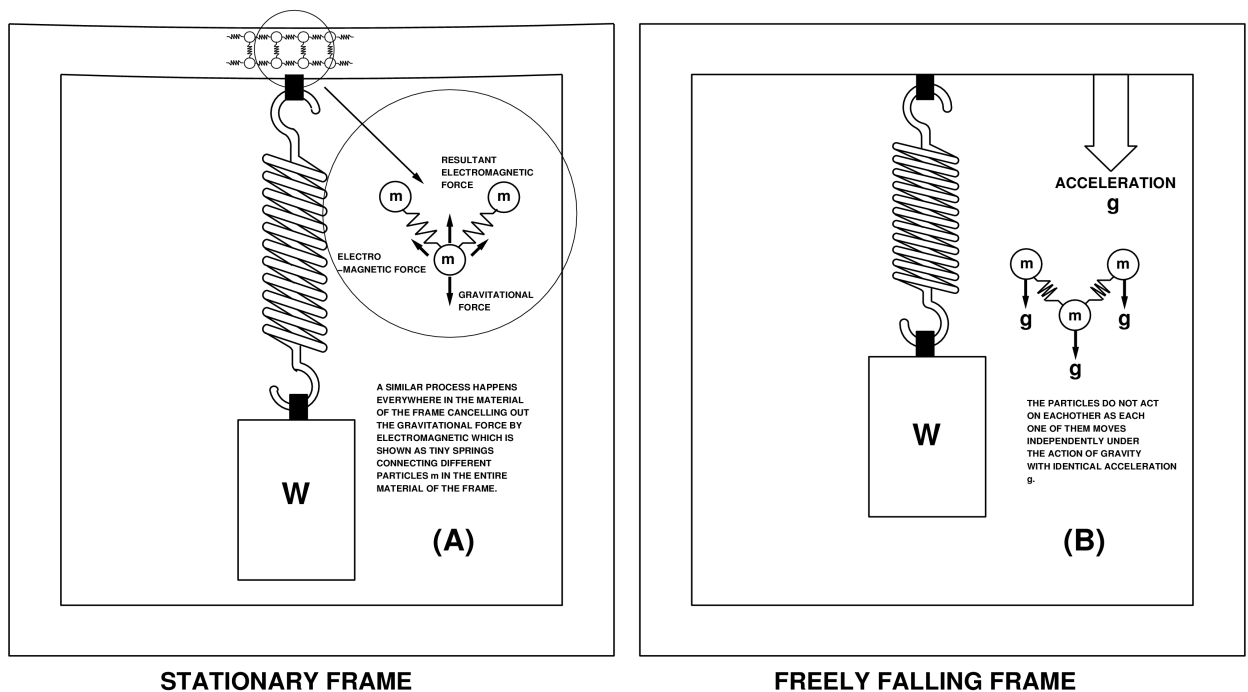

*Fig 1 : Illustration showing the difference between a stationary frame(A) and a freely falling frame(B).*

weight(which is the pull of gravity) the spring is stretched proportionally to the mass of the object. In a given spring balance if a mass of 1kg produces a stretch of 1cm then a mass of 2 kg produces a stretch of 2cm, assuming the spring exhibits linearity in stretch(Strelkov 1987) proportional to the force acting on the hook. This happens because a mass of 2kg is attracted by earth by a force that is twice as much as for the 1kg object. So the stretch is a measure of weight of the object. If an object did not have weight then it will not stretch the spring. Let us consider what happens in a frame consisting of a weight W suspended from a spring when it is stationary on earth and when it is falling freely towards earth. In the first case as shown in Figure 1A we see that all the parts of the frame are stationary with respect to earth. The force acting on the weight produces acceleration in it increasing its velocity. As a consequence the weight is displaced from its position towards earth. However its displacement stretches the spring which now begins to exert force on the weight that would accelerate it away from the earth(the spring pulls up the weight). At a certain point both the forces balance eachother and the weight comes to rest. It should be noted that in the stationary frame all the parts(any tiny particle of the material of the frame or spring) of the frame experience the force of gravity. As the test weight comes to rest after a certain displacement these parts also come to rest similarly. But their displacement is very small compared to that of the spring. We can think of the different parts of the frame being connected by tiny electromagnetic springs. The forces in the spring are also electromagnetic however due to its structure we see a larger stretch which is a result of addition of displacement of different parts of the spring. Most importantly we have to observe that in the stationary frame there are forces(electromagnetic against gravitational and viceversa) in balance between different parts of the body. This results in a person stationary on earth experiencing what we call weight. Next we consider the freely falling frame as in Figure 1B, here all the parts(including the spring) of the frame move freely or are accelerated identically under the force of gravity. As such the entire frame with everything within it can be thought of as one object which is moving freely with gravitational acceleration g. Most importantly the balance we had in the stationary frame between electromagnetic and gravitational forces does not arise in the freely falling frame. So the test weight does not exert a force on the spring hook as all the parts or particles of the frame accelerate identically throughout their fall and hence do not act upon eachother. On the contrary in the stationary frame inspite of the force of gravity acting on the particles ultimately all the particles in the frame are at rest. This can only happen when the net force is zero. This is brought about by another force, the electromagnetic force, between the particles of the entire material of the frame. Which arises when there is a relative displacement between material particles of the frame and causes them to move against the direction in which the gravitational force would. To show this we can build an apparatus that indicates loss of weight by objects falling freely under the action of gravity. In other words a spring balance is setup that closes a switch when the test object loses its weight. This switch in turn closes the circuit powering an LED. The glow of LED indicates that the object has lost its weight. The schematic circuit diagram of this apparatus is as shown in Figure 2. It is a simple circuit consisting of a battey(CR2032, 3.0V), an LED and a switch controlled by the weight of the test object W.

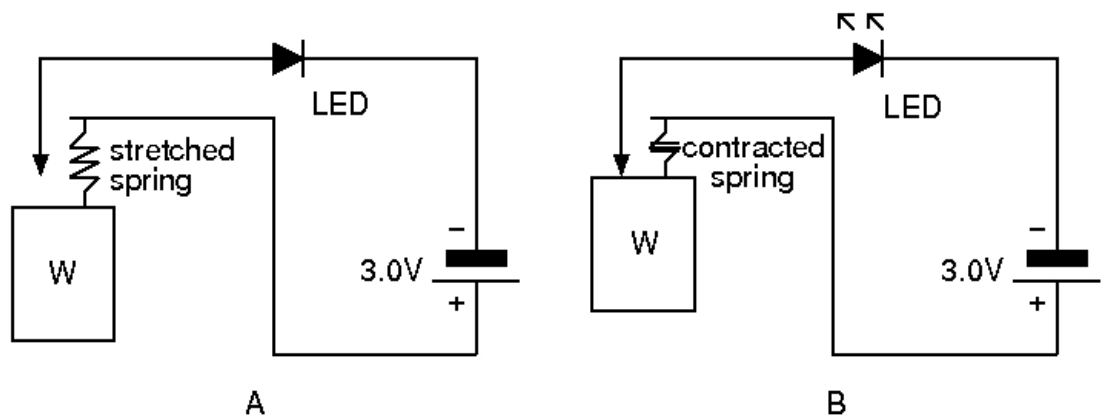

*Fig 2: Schematic circuit diagram to demonstrate free fall weightlessness. (A) is a rigid frame on earth in which the test object W has weight and hence it stretches the spring. (B) is a freely falling frame in which the test object loses its weight and hence the spring contracts closing the circuit that lights the LED.*

## 2. THE FREE FALL FRAME

The construction of the frame is shown in Figure 3. It should be noted that while it is possible to demonstrate free-fall weightlessness by merely throwing the frame vertically it is recommended to have 3 wires or rods to guide the free fall of the frame. This will protect the frame from damage at the same time improving the demonstration. The frame consists of a chassis which is a C-shaped double metal bracket. This can be constructed from 0.3mm-1mm thick metal(Al or Fe) sheet. The frame can be built of any size according to ones conveneince. Larger the frame, thicker the metal sheet required. On the frame two pillars/rods are provided which are separated by some distance(say 2-5cm). These guide the suspension of the weight and hold it in proper place during free-fall. The weight can be a suitable object(suggestion: iron core of a loud speaker) which can be suspended using a light spring like that from a writing pen. It should be ensured that the spring is stretched sufficiently(say atleast 1-3mm) under the action of weight when the frame is stationary. The contact bolt is fixed appropriately from the top of the frame such that it clearly makes a contact with the weight when the spring contracts fully assuming no weight was suspended from it. The LED and the battery are fixed on the top of the frame as shown in Figure 3/4. The wooden/plastic block that is fixed to the back of the frame has two holes through which the free-fall guiding rods/wires pass. Another free-fall guiding hole is provided in the front portion of the frame. Further another wooden/plastic block with a hole can be fixed to this so that the frame falls freely without wiggling. If desired metal bushing can be provided to these holes to prolong the life of the apparatus. To avoid damage to the frame, rubber/sponge pads on the lower side of the frame should be provided that absorb the ground impact shock. If required a resistor(47-100Ω) can be included in series to limit the LED current. Other possibilities are to use a white LED, however it seems red is a better choice. Figure 4 gives an illustration of the Free-Fall Frame.

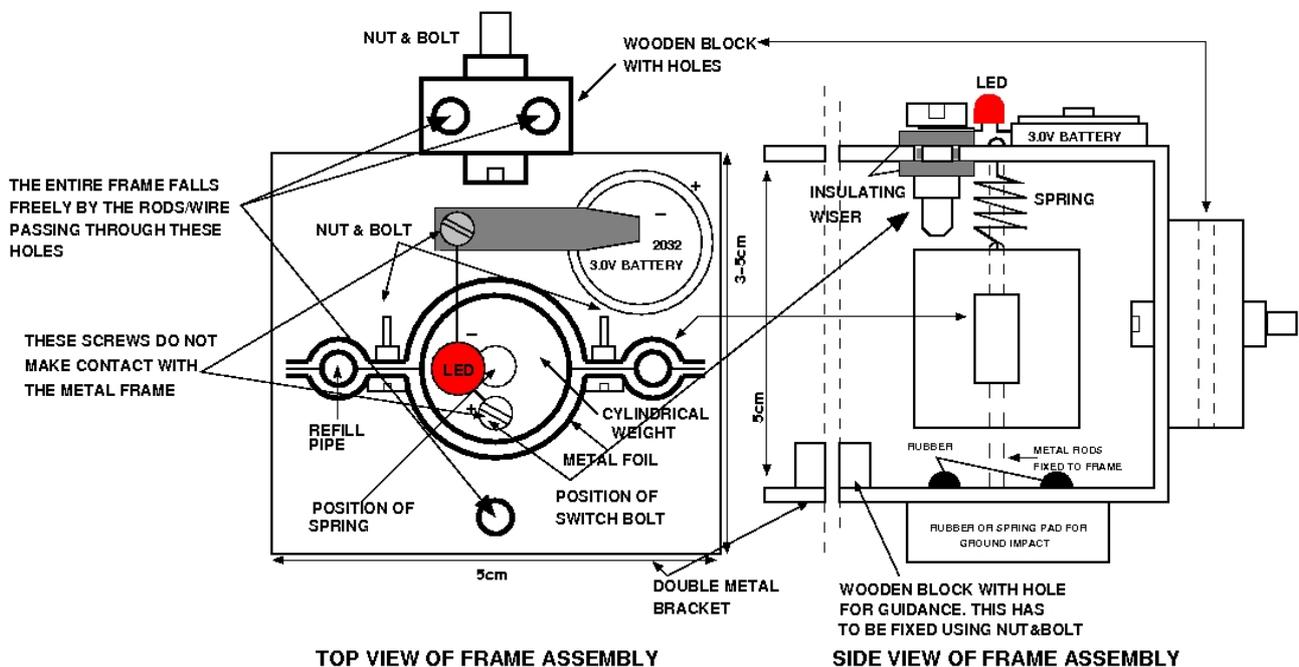

*Fig 3: Construction details of the Free-fall Frame. The frame can be built of any size according to available resources. A smaller frame is easily built using commonly available items. Also see Figure 4.*

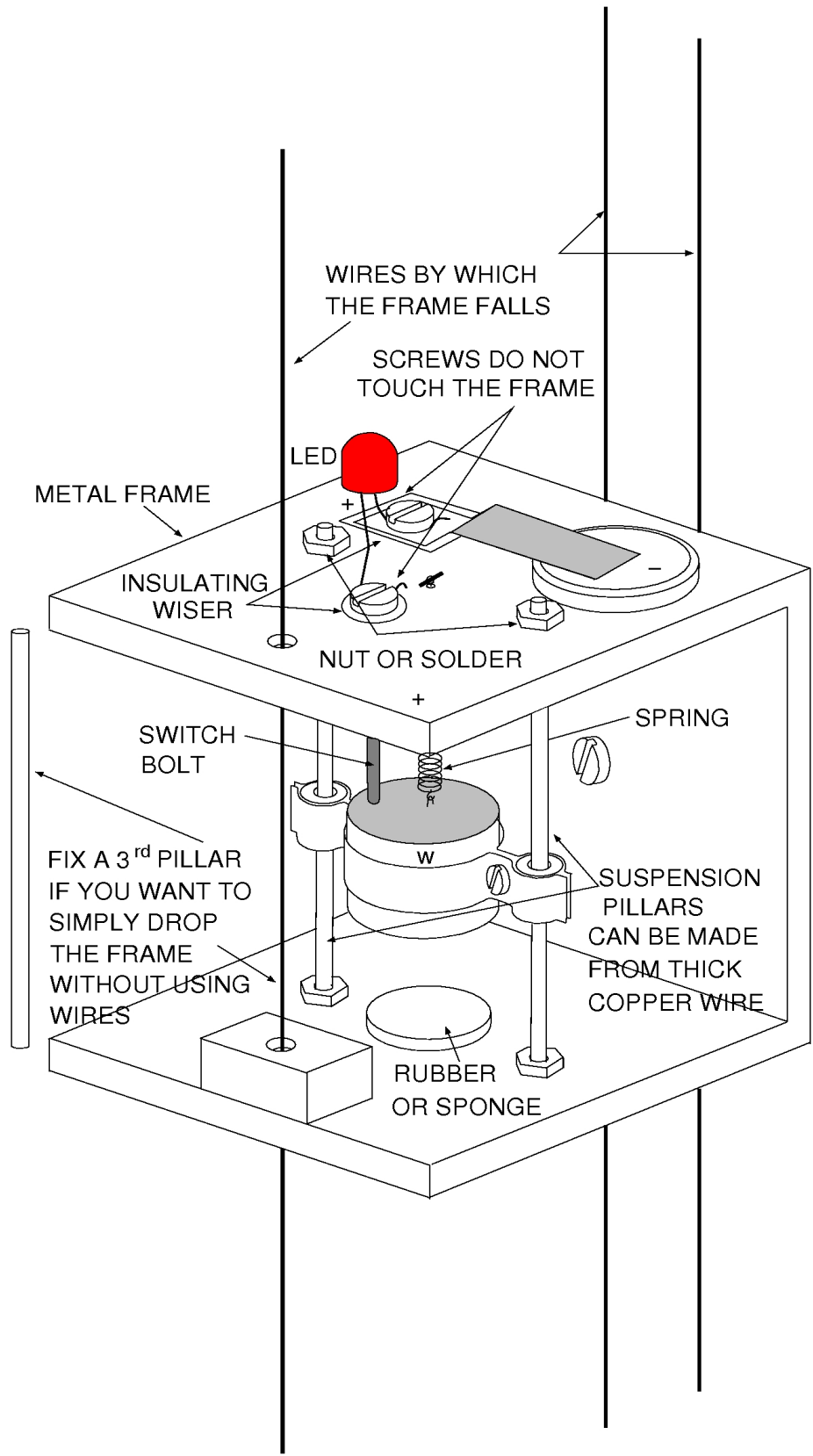

*Fig 4: Illustration showing the Free Fall Frame. Illustrations and diagrams in this document created using **Linux: xfig**.*

### 3. SUMMARY

This article discusses the weightlessness during free fall. The demonstration hinges on contraction of a spring. However it should be noted that weightlessness persists in the entire frame which falls as one object under gravitational acceleration. The spring being a highly elastic structure in the frame producing significant elongation upon the action of a force and being connected between a massive object like W and the frame roof serves to exhibit the interaction of different parts in the frame. Such an interaction exists between all the particles in the frame. This causes the the deformation of the frame roof as shown in Figure 1A. The difference is in the extent of displacement. The structure of the spring gives it the property of producing larger elongation(displacement) for a given force. This makes it suitable as one object in the frame to exhibit the interaction between different parts.